\documentclass[11pt,a4paper]{article}
\pdfoutput=1

\usepackage{jheppub}
\usepackage{amsmath,amsfonts,mathrsfs,graphicx,bbm}
\usepackage{fixltx2e}

\usepackage[font=small]{caption}

\usepackage{lmodern}

\usepackage{graphicx}
\usepackage{booktabs}
\usepackage{dsfont}

\usepackage{mathrsfs}
\usepackage{amsmath,amssymb}
\usepackage{mathtools}
\usepackage{bbm}
\usepackage{slashed}
\usepackage{braket}

\usepackage{hyperref}

\usepackage{xspace}

\usepackage{color}

\definecolor{grey}{rgb}{0.4,0.4,0.5}
\definecolor{darkgreen}{rgb}{0,0.5,0}
\definecolor{darkred}{rgb}{0.6,0.0,0}
\definecolor{lightbrown}{rgb}{1,0.9,0.8}
\definecolor{brown}{rgb}{0.6,0.3,0.3}
\definecolor{darkblue}{rgb}{0,0,0.8}
\definecolor{darkmagenta}{rgb}{0.5,0,0.5}

\def\cT{{\mathcal T}}
\def\de{\delta }

\numberwithin{equation}{section}

\makeatletter
 \let\old@startsection=\@startsection
 \let\oldl@section=\l@section
 \renewcommand{\@startsection}[6]{\old@startsection{#1}{#2}{#3}{#4}{#5}{#6\mathversion{bold}}}
 \renewcommand{\l@section}[2]{\oldl@section{\mathversion{bold}#1}{#2}}
\makeatother

\def\XXint#1#2#3{{\setbox0=\hbox{$#1{#2#3}{\int}$}
    \vcenter{\hbox{$#2#3$}}\kern-.5\wd0}}

\newcommand{\AdS}{\text{AdS}}

\newcommand{\alg}[1]{\mathfrak{#1}}

\def\be{\begin{equation}}
\def\ee{\end{equation}}

\newcommand{\bea}{\begin{eqnarray}}
\newcommand{\eea}{\end{eqnarray}}
\newcommand{\bei}{\begin{itemize}}
\newcommand{\eei}{\end{itemize}}
\newcommand{\bee}{\begin{enumerate}}
\newcommand{\eee}{\end{enumerate}}

\newcommand{\ads}{${\rm  AdS}_5\times {\rm S}^5\ $}

\def\ov{\over}

\def\la{\label}

\def\a {\alpha}
\def\b {\beta}

\def\g {\gamma}
\def\om {\omega}

\def\p{\phi}
\def\vp{\varphi}
\def\vk{\varkappa}
\def\e{\eta}
\def\t{\theta}
\def\z{\zeta}
\def\eps{\epsilon}
\def\r {\rho}

\def\S{\Sigma}

\def\pa {\partial}

\newcommand{\su}{\alg{su}}
\newcommand{\so}{\alg{so}}

\newcommand{\psu}{\alg{psu}}

\newcommand{\ag}{\alg{g}}

\def\L{\mathscr L}
\def\mI{\mathbbm{1}}

\def\qdp{\upsilon}
\def\dpnu{\nu}
\def\dpo{\eta}

\title{S-matrix for strings on $\eta$-deformed \ads }

\author[a,1]{Gleb Arutyunov,}
\note{Correspondent fellow at Steklov
Mathematical Institute, Moscow.}
\author[a]{Riccardo Borsato}
\author[b,c,1,2]{and Sergey Frolov}\note{On leave from Trinity College Dublin.}
\affiliation[a]{Institute for Theoretical
Physics and Spinoza Institute, Utrecht University, \\ Leuvenlaan \ 4, 3584 CE
Utrecht, The Netherlands}
\affiliation[b]{ Institut f\"ur Mathematik und Institut f\"ur Physik, Humboldt-Universit\"at zu Berlin, \phantom{$^\S$}\\
IRIS Adlershof, Zum Gro{\ss}en Windkanal 6, 12489 Berlin, Germany}
\affiliation[c]{Hamilton Mathematics Institute and School of Mathematics, \\
~~Trinity College, Dublin 2, Ireland}
\emailAdd{G.E.Arutyunov@uu.nl, R.Borsato@uu.nl, frolovs@maths.tcd.ie}

\abstract{We determine the bosonic part of the superstring sigma model Lagrangian on $\eta$-deformed AdS$_5\times$S$^5$, and use it to compute the perturbative world-sheet scattering matrix of bosonic particles of the model. We then compare it with the large string tension limit of the q-deformed S-matrix and find exact agreement. 

}

\begin{document}

\begin{flushright}\small{ITP-UU-13-31\\SPIN-13-23
\\
HU-Mathematik-2013-24\\
TCD-MATH-13-16}\end{flushright}

\maketitle
\flushbottom

\section{Introduction}
In recent years significant progress has been made towards understanding the excitation spectrum of strings moving in five-dimensional anti-de Sitter space-time and, accordingly,
the spectrum of scaling dimensions of composite operators in planar ${\cal N} = 4$ supersymmetric gauge theory. This progress became possible due to the fundamental insight that strings propagating in 
AdS space can be described by an integrable model.
In certain aspects, however, the deep origin of this exact solvability has not yet been  unraveled, mainly because of 
tremendous complexity of the corresponding model.  A related question concerns robustness of integrability in the context of the gauge-string correspondence \cite{M},
as well as the relationship between integrability and the amount of global (super)symmetries preserved by the target space-time in which strings propagate. 
To shed further light on these important issues, 
one may attempt to search for new examples of integrable string backgrounds that can be solved by similar techniques. One such instance, where this program is largely promising to succeed, is to study various deformations of the string target space that preserve the integrability of the two-dimensional quantum field theory on the world sheet. Simultaneously, this should provide interesting new information about integrable string models and their dual gauge theories.

\smallskip

There are two known classes of integrable deformations of the \ads superstring. The first of these is a class of backgrounds obtained either by orbifolding \ads by a discrete subgroup of the corresponding isometry group 
\cite{Kachru:1998ys,Lawrence:1998ja}  or  
by applying a sequence of T-duality -- shift -- T-duality transformations (also known as $\gamma$-deformations) to this space, giving a string theory on a TsT-transformed background \cite{LM,F05}. 
Eventually all deformations of this class can be conveniently described in terms of the original string theory, where the deformations result into quasi-periodic but still integrable 
boundary conditions for the world-sheet fields.

\smallskip

The second class of deformations affects the \ads model on a much more fundamental level and is related to deformations of the underlying symmetry algebra.
In the light-cone gauge this symmetry algebra constitutes two copies of the centrally extended  Lie superalgebra $\psu(2|2)$ with the same central extension for each copy.
It appears that this centrally extended $\psu(2|2)$, or more precisely its universal enveloping algebra,
admits a natural deformation $\psu_q(2|2)$ in the sense of quantum groups \cite{Beisert:2008tw,Beisert:2011wq}. This
algebraic structure is the starting point for the construction of a $\psu_q(2|2)\oplus \psu_q(2|2)$-invariant S-matrix, giving
a quantum deformation of the \ads world-sheet S-matrix \cite{Beisert:2008tw, Ben,deLeeuw:2011jr}. The deformation parameter
$q$ can be an arbitrary complex number, but  in
physical applications is typically taken to be either real or a root of unity.   

\smallskip

Since these quantum group deformations modify the dispersion relation and the scattering matrix, to solve the corresponding model by means of the mirror Thermodynamic Bethe Ansatz (TBA), for a recent review see \cite{Stijn}, one has to go through the entire procedure of first deriving the TBA equations for the ground state and then extending them to include excited states. While this program has been successfully carried out for deformations with $q$ being a root of unity \cite{Arutyunov:2012zt,Arutyunov:2012ai}, the corresponding string background remains unknown. There is a conjecture that  in the limit of infinite 't Hooft coupling the q-deformed S-matrix tends to that of the Pohlmeyer-reduced version of the \ads superstring \cite{Beisert:2010kk,Hoare:2013ysa}. It is not straightforward, however, to identify the S-matrix of the latter theory as one has to understand whether the elementary excitations that scatter in that model are solitons or kinks.

\smallskip 

The case of real deformation parameter considered in this paper is not less compelling. Recently there was an interesting proposal on how to deform the sigma-model for strings on \ads with a real deformation parameter  
$\dpo$ \cite{Delduc:2013qra}.
 Deformations of this type constitute a general class of deformations
 governed by solutions of the classical Yang-Baxter equation   \cite{Cherednik:1981df,Klimcik:2008eq}. This class is not solely restricted to the string model in question but can be applied to a 
 large variety of two-dimensional integrable models based on (super)groups or their cosets  \cite{Klimcik:2002zj}-\cite{Kawaguchi:2012gp}.

\smallskip

The aim of the present work is to compute the $2\to 2$ scattering matrix for the $\dpo$-deformed model in the limit of large string tension $g$ and to compare the corresponding result with the known q-deformed S-matrix found from quantum group symmetries, unitarity and crossing \cite{Beisert:2008tw, Ben}.  In the context of the undeformed model a computation of this type has been carried out in \cite{Klose:2006zd}.

\smallskip

The $\dpo$-deformed model appears to be rather involved, primarily because of fermionic degrees of freedom. 
Our strategy is therefore to switch off fermions and proceed by studying the corresponding bosonic action.   
Of course, the perturbative S-matrix computed from this action will not coincide with the full world-sheet S-matrix but nevertheless
will give a sufficient part of the scattering data to provide a non-trivial test for both integrability (the Yang-Baxter equation) and a comparison 
with the q-deformed S-matrix. The contribution of fermions is of secondary concern and will be discussed elsewhere \cite{ABF}.
 
 \smallskip
Let us summarize the results of this paper. Coming back to the bosonic action, we find that it corresponds to a string background which in addition to the metric also supports a non-vanishing $B$-field. 
 The deformation breaks \ads isometries down to ${\rm U}(1)^3\times {\rm U}(1)^3$, where the first and second factors refer to the deformed AdS and five-sphere, respectively.   
 Thus, only isometries corresponding to the Cartan elements of the isometry algebra of the \ads survive, very similar to the case of generic $\gamma$-deformations. 
 As for the metric, its AdS part exhibits a singularity whose nature is currently unclear. Computed in a string frame the metric  includes the contribution of a dilaton, and
 to extract the latter one needs to know the RR-fields which requires considering fermions.    

\smallskip 

With the bosonic action at hand it is straightforward to compute the corresponding tree-level S-matrix. We then show that  it matches
perfectly  with the q-deformed S-matrix taken in the large tension limit and restricted to the scattering of bosons, provided we identify the deformation parameters as 
$$
q=e^{-\nu/g}\, ,\quad \dpnu = {2\e\ov 1+\e^2}\,.
$$
This is the main result of our work which makes it quite credible that the $\dpo$-deformed model indeed may enjoy hidden $\psu_q(2|2)\oplus \psu_q(2|2)$ symmetry for finite 
values of the coupling constant $g$. If true, this implies that despite the singular behaviour of the metric the quantum string sigma model would be well defined. In particular it would be possible to compute its exact spectrum by means of the mirror TBA.

\smallskip
The paper is organized as follows. In the next section we recall the general form of the action for the $\dpo$-deformed model and use it to derive an explicit form of the Lagrangian for bosonic 
degrees of freedom.  In section 3, upon fixing the uniform light-cone gauge, the corresponding Hamiltonian is derived up to quartic order in fields  and further used to compute 
the tree-level S-matrix. This result is subsequently compared to the one arising from the q-deformed S-matrix (which includes the dressing phase) in the large $g$ limit. We conclude by outlining
interesting open problems. Finally some technical details on the derivation of the bosonic Lagrangian, the perturbative expansion of the q-deformed dressing phase and the form of the q-deformed 
S-matrix are collected in three appendices.

\section{Superstrings on $\eta$-deformed \ads}
According to \cite{Delduc:2013qra}, the action for superstrings on the deformed \ads is 
\bea\nonumber
S=\int {\rm d\sigma}{\rm d\tau} \mathscr{L}\, ,
\eea
where the Lagrangian density depending on a real deformation parameter $\eta$ is given by\footnote{Note that our $\e$-dependent prefactor  differs from the one in \cite{Delduc:2013qra}. Our choice is necessary to match the perturbative world-sheet scattering matrix with the q-deformed one.}
\bea
\label{defLag}
{\mathscr L}=-\frac{g}{4}(1+\eta^2)\big(\gamma^{\a\b}-\eps^{\a\b}\big)\, {\rm str}\Big[\tilde{d}(A_{\a})\frac{1}{1-\eta R_\ag \circ d}(A_{\beta}) \Big]\, .
\eea
Here and in what follows we use the notations and conventions from  \cite{Arutyunov:2009ga}, in particular $\eps^{\tau\sigma}=1$ and $\gamma^{\a\b}=h^{\a\b}\sqrt{-h}$ that is the Weyl invariant combination of the world-sheet metric $h_{\a\b}$; the component $\gamma^{\tau\tau}<0$.
The coupling constant $g$ is the effective string tension. 
Further, $A_{\a}=-\ag^{-1}\pa_{\a}\ag$, where $\ag\equiv \ag(\tau,\sigma)$ is a  coset representative from ${\rm PSU}(2,2|4)/{\rm SO}(4,1)\times {\rm SO}(5)$. 
To define the operators
$d$ and $\tilde{d}$ acting on the currents $A_{\alpha}$, we need to recall that the Lie superalgebra $\mathscr{G}=\psu(2,2|4)$ admits a ${\mathbb Z}_4$-graded decomposition 
$$
\mathscr{G}=\mathscr{G}^{(0)}\oplus \mathscr{G}^{(1)}\oplus \mathscr{G}^{(2)}\oplus \mathscr{G}^{(3)}\, .$$
Here $\mathscr{G}^{(0)}$ coincides with $\so(4,1)\times \so(5)$. Denoting by $P_i$, $i=0,1,2,3$, projections on the corresponding components of the graded decomposition above, 
operators $d$ and $\tilde{d}$ are defined as 
\bea
\nonumber
d&=&P_1+\frac{2}{1-\eta^2}P_2-P_3,\\
\nonumber
\tilde{d}&=&-P_1+\frac{2}{1-\eta^2}P_2+P_3\, .
\eea
Finally, the action of the operator $R_\ag$ on $M\in {\mathscr G}$ is given by  
\be\label{Rgop}
R_\ag(M) = \ag^{-1}R(\ag M\ag^{-1})\ag\, ,
\ee
where $R$ is a linear operator on $\mathscr{G}$ satisfying the modified classical Yang-Baxter equation. In the following we define the action of $R$ on an arbitrary $8\times 8$ matrix $M$ 
as 
\be\label{Rop}
R(M)_{ij} = -i\, \eps_{ij} M_{ij}\,,\quad \eps_{ij} = \left\{\begin{array}{ccc} 1& \rm if & i<j \\
0&\rm if& i=j \\
-1 &\rm if& i>j \end{array} \right.\,,
\ee
In the limit $\eta\to 0$ one recovers from  (\ref{defLag}) the Lagrangian density of the \ads superstring. 

\smallskip

Our goal now is to obtain an explicit form for the corresponding bosonic action.
With fermionic degrees of freedom switched off, formula (\ref{defLag}) simplifies to 
\bea
\label{defLagbos}
{\mathscr L}=- \frac{g}{2}(1+\varkappa^2)^{1\ov2}\, \big(\gamma^{\a\b}-\eps^{\a\b}\big){\rm str}\, \Big[A_{\a}^{(2)}\frac{1}{1-\varkappa R_\ag \circ P_2}(A_{\beta}) \Big]\, ,
\eea
where we have introduced 
$$
\varkappa=\frac{2\eta}{1-\eta^2}\,,
$$
which as we see in a moment is a convenient deformation parameter.

To proceed, we need to choose a representative $\ag$ of a bosonic coset ${\rm SU}(2,2|4)\times {\rm SU}(4)/{\rm SO}(4,1)\times {\rm SO}(5)$ and invert the operator 
$1-\eta R_\ag \circ d$. 
A convenient choice of a coset representative and the inverse of  $1-\eta R_\ag \circ d$ are discussed in appendix \ref{IRgd}. Making use of the inverse operator, one can easily compute the corresponding bosonic Lagrangian. 
It is given by the sum of the AdS and sphere parts
\be\la{Lfull}
\L = \L_{\alg{a}} +\L_{\alg{s}}= \L_{\alg{a}}^{G} +\L_{\alg{a}}^{WZ} +\L_{\alg{s}}^{G} +\L_{\alg{s}}^{WZ}\, ,
\ee
where we further split each part into the contribution of the metric and Wess-Zumino pieces. Accordingly, for the metric pieces we obtain 
  \bea\nonumber
\L_{\alg{a}}^{G} &=&-{g\ov2}(1+\varkappa^2)^{1\ov2}\, \g^{\a\b}\Big(
-\frac{\pa_\a t\pa_\b t\left(1+\rho ^2\right)}{ 1-\varkappa ^2 \rho ^2}
+\frac{\pa_\a \r\pa_\b \r}{ \left(1+\rho ^2\right) \left(1-\varkappa ^2 \rho ^2\right)}
+\frac{\pa_\a \z\pa_\b\z \rho ^2}{1+ \varkappa ^2 \rho ^4 \sin ^2\z }
\\\la{LaG}
   &&\qquad\qquad\qquad+\frac{\pa_\a \psi_1\pa_\b\psi_1\rho ^2 \cos
   ^2\z}{ 1+\varkappa ^2 \rho ^4 \sin ^2\z}+\pa_\a \psi_2\pa_\b\psi_2
  \rho ^2 \sin ^2\z \Big)\,,
\eea
\bea\nonumber
\L_{\alg{s}}^{G} &=&-{g\ov2}(1+\varkappa^2)^{1\ov2}\, \g^{\a\b}\Big(\frac{\pa_\a \p\pa_\b \p
  \left(1-r^2\right)}{1+\varkappa ^2 r^2}+\frac{\pa_\a r\pa_\b r
  }{ \left(1-r^2\right) \left(1+\varkappa ^2 r^2\right)}
  +\frac{\pa_\a \xi\pa_\b \xi  r^2}{1+ \varkappa ^2 r^4 \sin ^2\xi}
  \\\la{LsG}
   &&\qquad\qquad\qquad+\frac{\pa_\a \p_1\pa_\b \p_1  r^2 \cos ^2\xi }{1+ \varkappa ^2
   r^4 \sin ^2\xi } +\pa_\a \p_2\pa_\b \p_2  r^2 \sin^2\xi \Big)\,,
\eea
while the Wess-Zumino parts $\L_{\alg{a}}^{WZ}$ and $\L_{\alg{s}}^{WZ}$ read
\bea\la{LaWZ}
\L_{\alg{a}}^{WZ} &=&{g\ov2} \varkappa (1+\varkappa^2)^{1\ov2}\, \eps^{\a\b}\frac{ \rho ^4 \sin 2 \zeta}{1+ \varkappa ^2 \rho ^4 \sin ^2\z}\pa_\a\psi_1\pa_\b\zeta\,,
\eea
\bea\la{LsWZ}
\L_{\alg{s}}^{WZ} &=&-{g\ov2} \varkappa (1+\varkappa^2)^{1\ov2}\, \eps^{\a\b}\frac{ r^4 \sin 2 \xi }{1+ \varkappa ^2 r^4 \sin^2\xi}\pa_\a\p_1\pa_\b\xi\,  .
\eea
Here the coordinates $t\,,\,\psi_1\,,\,\psi_2\,,\, \z\,,\, \r$ parametrize the deformed AdS space, while the coordinates $\phi\,,\,\p_1\,,\,\p_2\,,\, \xi\,,\, r$ parametrize the deformed five-sphere. Switching off the deformation, one finds that
the AdS$_5$ coordinates are related to the embedding 
coordinates $Z^A,$ $A = 0,1,\ldots,5$ obeying the constraint $\eta^{AB}Z_AZ_B=-1$ where $\eta^{AB}=(-1,1,1,1,1,-1)$ as
\bea
Z_1+iZ_2 = \r\cos\z\,e^{i\psi_1}\,,\quad Z_3+iZ_4 = \r\sin\z\,e^{i\psi_2}\,,\quad Z_0+iZ_5 = \sqrt{1+\r^2}\,e^{it}\,,
\eea
while the $S_5$ coordinates  are related to the embedding 
coordinates $Y^A,$ $A = 1,\ldots,6$ obeying $Y_A^2=1$  as
\bea
Y_1+iY_2 = r\cos\xi\,e^{i\p_1}\,,\quad Y_3+iY_4 = r\sin\xi\,e^{i\p_2}\,,\quad Y_5+iY_6 = \sqrt{1-r^2}\,e^{i\p}\,.
\eea
It is obvious  that the deformed action is invariant under ${\rm U}(1)^3\times {\rm U}(1)^{3}$ corresponding to the shifts of $t$, $\psi_k$, $\p$, $\p_k$. One also finds the ranges of $\r$ and $r$: $0\le \r\le{1\ov\vk}$ and $0\le r\le1$.
The (string frame) metric of the deformed AdS is singular at $\r=1/\vk$. Since we do not know the dilaton it is unclear if the Einstein frame metric exhibits the same singularity.  The bosonic Wess-Zumino terms signify the presence of a non-trivial background $B$-field which is absent in the undeformed case. 

In the next section we are going to impose the light-cone gauge, take the decompactification limit and compute the bosonic part of the four-particle world-sheet scattering matrix. To this end, we first expand the Lagrangian \eqref{Lfull} up to quartic order in $\r$, $r$ and their derivatives, and then make the shifts of $\rho$ and $r$ as described in appendix \ref{IRgd}, {\it c.f.}  (\ref{shift}).
Since we are interested in the perturbative expansion in powers of fields around $\rho=0$, 
the final step consists in changing the spherical coordinates to $(z_i,y_i)_{i=1,\ldots,4}$ as
\bea
\begin{aligned}
\label{flatcoord}
\frac{z_1+iz_2}{1-\tfrac{1}{4}z^2}=\rho\cos\zeta e^{i\psi_1}\, , ~~~~~~~\frac{z_3+iz_4}{1-\tfrac{1}{4}z^2}=\rho\sin\zeta e^{i\psi_2} \, , ~~~~~z^2\equiv z_i^2\, ,\\
\frac{y_1+iy_2}{1+\tfrac{1}{4}y^2}=r\cos\xi e^{i\phi_1}\, , ~~~~~~~\frac{y_3+iy_4}{1+\tfrac{1}{4}y^2}=r\sin\xi e^{i\phi_2} \, ,~~~~~~y^2\equiv y_i^2 \, ,
\end{aligned}
\eea
with further expanding the resulting action up to the quartic order in $z$ and $y$ fields.  In this way we find the following quartic Lagrangian
\begin{equation}\la{Lquart}
\begin{aligned}
\L_{\alg{a}} &= -\frac{g}{2} (1+\varkappa^2)^{1\ov2} \,  \gamma^{\alpha \beta} \Bigg[ -\left( 1 + (1+\varkappa^2) z^2 +\frac{1}{2}(1+\varkappa^2)^2(z^2)^2\right) \pa_{\a}t \pa_{\b}t \\
& + \left(1+(1-\varkappa^2)\frac{z^2}{2}\right) \pa_{\a}z_i\pa_{\beta}z_i \Bigg]+2g \varkappa(1+\varkappa^2)^{1\ov2} (z_3^2+z_4^2) \epsilon^{\a\b} \pa_{\a}z_1 \pa_{\b}z_2 \, , \\
\L_{\alg{s}}&= -\frac{g}{2} (1+\varkappa^2)^{1\ov2}\,  \gamma^{\alpha \beta} \Bigg[ \left( 1 - (1+\varkappa^2) y^2 +\frac{1}{2}(1+\varkappa^2)^2(y^2)^2\right) \pa_{\a}\phi \pa_{\b}\phi \\
& + \left(1-\frac{1}{2}(1-\varkappa^2)y^2 \right) \pa_{\a}y_i\pa_{\beta}y_i  \Bigg] - 2 g\varkappa(1+\varkappa^2)^{1\ov2} (y_3^2+y_4^2) \epsilon^{\a\b} \pa_{\a}y_1 \pa_{\b}y_2\, .
\end{aligned}
\end{equation}
We point out that the metric part of this Lagrangian has a manifest ${\rm SO}(4)\times {\rm SO(4)}$ symmetry which is however broken by the Wess-Zumino terms.

\section{Perturbative bosonic world-sheet S-matrix}

\subsection{Light-cone gauge and quartic Hamiltonian}

To fix the light-cone gauge and compute the scattering matrix, it is advantageous to use the Hamiltonian formalism.  For the reader's convenience we start with a general discussion on how to construct the Hamiltonian for the world-sheet action 
of the form
\begin{equation}
S=-\frac{g}{2} \int_{-r}^r \, {\rm d}\sigma {\rm d} \tau \left( \, \gamma^{\alpha\beta} \partial_\alpha X^M \partial_\beta X^N G_{MN} -\epsilon^{\alpha\beta} \partial_\alpha X^M \partial_\beta X^N B_{MN} \right),
\end{equation}
where $G_{MN}$ and $B_{MN}$ are the background metric and $B$-field respectively.
In the first order formalism we introduce conjugate momenta
\begin{equation}
p_M = \frac{\delta S}{\delta \dot{X}^M} = - g \gamma^{0\beta} \partial_\beta X^N G_{MN} + g X^{'N} B_{MN}.
\end{equation}
The action can be rewritten as
\begin{equation}
S=  \int_{-r}^r \, {\rm d}\sigma {\rm d} \tau \left( p_M \dot{X}^M + \frac{\gamma^{01}}{\gamma^{00}} C_1 + \frac{1}{2g \gamma^{00}} C_2  \right),
\end{equation}
where $C_1, C_2$ are the Virasoro constraints.
They are given by
\bea
C_1 &=& p_M X'^{M}, \\ 
C_2 &=& G^{MN} p_M p_N - 2 g p_M X'^{Q} G^{MN} B_{NQ} + g^2 X'^{P} X'^{Q} B_{MP} B_{NQ} G^{MN} + g^2 X'^{M} X'^{N} G_{MN}. \nonumber
\eea
The first Virasoro constraint has the same form as in the undeformed case. In particular, the solution for $x_-'$ in terms of $p_\mu,x_\mu$ will still be the same.
When expressed in terms of the conjugate momenta, the second constraint gets an explicit dependence on the B-field.
To impose light-cone gauge, one first introduces light-cone coordinates
\begin{equation}\la{lcg}
x_-=\phi-t, \qquad x_+= (1-a)t +a \phi.
\end{equation}
The second Virasoro constraint can be written as
\begin{equation}
\begin{aligned}
C_2 &= G^{--} p_-^2 +2 G^{+-} p_+ p_- + G^{++} p_+^2 \\
& + g^2 G_{++} x_-'^2 + 2 g^2 G_{+-} x_+' x_-' + g^2 G_{--} x_+'^2 + \mathcal{H}_x\, ,
\end{aligned}
\end{equation}
where
\begin{equation}
\begin{aligned}
\nonumber
G^{--} &= a^2 G_{\phi\phi}^{-1} - (a-1)^2 G_{tt}^{-1}, \qquad &G^{+-}& = a G_{\phi\phi}^{-1} - (a-1) G_{tt}^{-1}, \qquad G^{++} = G_{\phi\phi}^{-1} - G_{tt}^{-1}, \\
G_{++} &= (a-1)^2 G_{\phi\phi} - a^2 G_{tt}, \qquad &G_{+-} &= -(a-1) G_{\phi\phi} + a G_{tt}, \qquad G_{--} = G_{\phi\phi} - G_{tt},
\end{aligned}
\end{equation}
and $\mathcal{H}_x$ is the part that depends on the transverse fields only
\begin{equation}
\mathcal{H}_x = G^{\mu\nu} p_\mu p_\nu + g^2 X'^\mu X'^\nu G_{\mu\nu} -2 g p_\mu X'^{\rho} G^{\mu\nu} B_{\nu\rho} + g^2 X'^\lambda X'^\rho B_{\mu\lambda} B_{\nu\rho} G^{\mu\nu}.
\end{equation}
Notice that the $B$-field is contained only in $\mathcal{H}_x$, since in the action it does not couple to the derivatives of $x_\pm$.
We impose the uniform light-cone gauge 
\begin{equation}
x_+= \tau, \qquad p_+=1.
\end{equation}
Solving $C_2=0$ for $p_-$ gives the Hamiltonian
\begin{equation}
\mathcal{H} = -p_-(p_\mu,x^\mu,x'^\mu).
\end{equation}
Formally the solution for the Hamiltonian is still given by eq. (2.16) of the review \cite{Arutyunov:2009ga}, with the only difference that now the components of the metric are deformed and that $\mathcal{H}_x$ has also the $B$-field contribution.
Rescaling the fields with powers of $g$ and expanding in $g$ one can find $\mathcal{H}_n$, namely the part of the Hamiltonian that is of order $n$ in the fields.
Then the action acquires the form
\begin{equation}
S= \int {\rm d}\tau {\rm d} \sigma \, \left( p_\mu \dot{x}^\mu - \mathcal{H}_2 - \frac{1}{g} \mathcal{H}_4 - \cdots \right),
\end{equation}
where the quadratic Hamiltonian is given by
\begin{equation}
\mathcal{H}_2 = \frac{1}{2} p_\mu^2 + \frac{1}{2} (1+\varkappa^2) x_\mu^2 + \frac{1}{2} (1+\varkappa^2) x'^2_\mu.
\end{equation}
The quartic Hamiltonian in a general $a$-gauge is 
\begin{equation}
\begin{aligned}
\mathcal{H}_4 &= \frac{1}{4} \Bigg( (2 \varkappa^2 z^2 -(1+\varkappa^2) y^2 ) p_z^2 - (2 \varkappa^2 y^2 -(1+\varkappa^2) z^2 ) p_y^2 \\
&+\left(1+\varkappa ^2\right) \left(\left(2 z^2-\left(1+\varkappa ^2\right) y^2\right)z'^2 + \left(\left(1+\varkappa ^2\right) z^2-2
   y^2\right)y'^2\right)\Bigg) \\
&- 2 \varkappa  \left(1+\varkappa ^2\right)^{1\ov2} \left(\left(z_3^2+z_4^2\right) \left(p_{z_1} z_2'-p_{z_2} z_1'\right) -  \left(y_3^2+y_4^2\right) \left(p_{y_1} y_2'-p_{y_2} y_1'\right) \right) \\
&+\frac{(2a-1)}{8} \Bigg( (p_y^2+p_z^2)^2 -(1+\varkappa^2)^2 (y^2+z^2)^2  \\
&+2 (1+\varkappa^2)(p_y^2+p_z^2)(y'^2+z'^2)+(1+\varkappa^2)^2 (y'^2+z'^2)^2 -4 (1+\varkappa^2) (x_-')^2\Bigg).
\end{aligned}
\end{equation}
Here we use the notation $p_z^2\equiv  p_{z_i}^2, \ p_y^2\equiv  p_{y_i}^2$, where sum over $i$ is assumed. 

To simplify the quartic piece, we  
can remove the terms of the form $p_z^2y^2$ and $p_y^2z^2$ by performing a canonical transformation generated by
\begin{equation}
V= \frac{(1+\varkappa^2)}{4} \int {\rm d}\sigma \Big( p_y y z^2 -p_z z y^2  \Big) ,
\end{equation}
where the shorthand notation $p_y y \equiv  p_{y_i} {y_i}, \  p_z z \equiv  p_{z_i} {z_i}$ was used. After this is done the quartic Hamiltonian is
\bea
\mathcal{H}_4 &=& \frac{(1+\varkappa^2)}{2} ( z^2 z'^2- y^2 y'^2 ) + \frac{(1+\varkappa^2)^{2}}{2} (z^2 y'^2-y^2 z'^2) +\frac{\varkappa^2}{2} ( z^2  p_z^2 - y^2  p_y^2 ) \nonumber \\
&-& 2\varkappa(1+\varkappa^2)^{1\ov2}  \left[\left(z_3^2+z_4^2\right) \left(p_{z_1} z_2'-p_{z_2} z_1'\right) -  \left(y_3^2+y_4^2\right) \left(p_{y_1} y_2'-p_{y_2} y_1'\right) \right] \nonumber \\
&+&\frac{(2a-1)}{8}\Bigg( (p_y^2+p_z^2)^2 -  (1+\varkappa^2)^2 (y^2+z^2)^2  \\
&+&2  (1+\varkappa^2) (p_y^2+p_z^2)(y'^2+z'^2)+ (1+\varkappa^2)^2 (y'^2+z'^2)^2 -4  (1+\varkappa^2) (x_-')^2\Bigg). \nonumber
\eea

We recall that in the undeformed case the corresponding theory is invariant with respect to the two copies of the centrally extended superalgebra $\psu(2|2)$, each containing two
$\su(2)$ subalgebras.  To render invariance under $\su(2)$ subalgebras manifest, one can introduce  two-index notation for the world-sheet fields. It is also convenient to adopt the same 
notation for the deformed case\footnote{This parameterisation is different from the one used in \cite{Arutyunov:2009ga} and the difference is the exchange of the definitions for $Y^{1\dot{1}}$ and $Y^{2\dot{2}}$. This does not matter in the undeformed case but is needed here in order to correctly match the perturbative S-matrix with the q-deformed one computed from symmetries.}
\begin{equation}
\begin{aligned}
&Z^{3\dot{4}} =\tfrac{1}{2} (z_3-i z_4),  \qquad &Z^{3\dot{3}} =\tfrac{1}{2} (z_1-i z_2), \\
& Z^{4\dot{3}}=-\tfrac{1}{2} (z_3+i z_4),    \qquad &Z^{4\dot{4}}=\tfrac{1}{2} (z_1+i z_2), 
\end{aligned}
\end{equation}   
\begin{equation}
\begin{aligned}
&Y^{1\dot{2}}=\tfrac{1}{2} (y_3-i y_4), \qquad   &Y^{1\dot{1}}=\tfrac{1}{2} (y_1+i y_2), \\
&Y^{2\dot{1}}=-\tfrac{1}{2} (y_3+i y_4), \qquad &Y^{2\dot{2}}=\tfrac{1}{2} (y_1-i y_2)\, .
\end{aligned}
\end{equation}
In terms of two-index fields the quartic Hamiltonian becomes $\mathcal{H}_4 = \mathcal{H}^G_4 + \mathcal{H}^{WZ}_4$,
where $\mathcal{H}^G_4$ is the contribution coming from the spacetime metric and $\mathcal{H}^{WZ}_4 $ from the $B$-field
{\small
\bea
\mathcal{H}^G_4 &=&2(1+\varkappa^2) \left( Z_{\alpha\dot{\alpha}} Z^{\alpha\dot{\alpha}} Z'_{\beta\dot{\beta}} Z'^{\beta\dot{\beta}} -Y_{a\dot{a}}Y^{a\dot{a}} Y'_{b\dot{b}}Y'^{b\dot{b}} \right) \nonumber \\
&+& 2(1+\varkappa^2)^{2} \left( Z_{\alpha\dot{\alpha}} Z^{\alpha\dot{\alpha}} Y'_{b\dot{b}}Y'^{b\dot{b}} - Y_{a\dot{a}}Y^{a\dot{a}} Z'_{\beta\dot{\beta}} Z'^{\beta\dot{\beta}} \right) \nonumber \\
& +&\frac{\varkappa^2}{2} \left( Z_{\alpha\dot{\alpha}} Z^{\alpha\dot{\alpha}} P_{\beta\dot{\beta}} P^{\beta\dot{\beta}} - Y_{a\dot{a}}Y^{a\dot{a}} P_{b\dot{b}}P^{b\dot{b}} \right) \nonumber \\
&+&\frac{(2a-1)}{8} \Bigg( \frac{1}{4}(P_{a\dot{a}}P^{a\dot{a}}+P_{\alpha\dot{\alpha}}P^{\alpha\dot{\alpha}})^2 -4 (1+\varkappa^2)^2 (Y_{a\dot{a}}Y^{a\dot{a}}+Z_{\alpha\dot{\alpha}}Z^{\alpha\dot{\alpha}})^2  \\
&+&2 (1+\varkappa^2) (P_{a\dot{a}}P^{a\dot{a}}+P_{\alpha\dot{\alpha}}P^{\alpha\dot{\alpha}})(Y'_{a\dot{a}}Y'^{a\dot{a}}+Z'_{\alpha\dot{\alpha}}Z'^{\alpha\dot{\alpha}})+4 (1+\varkappa^2)^2 (Y'_{a\dot{a}}Y'^{a\dot{a}}+Z'_{\alpha\dot{\alpha}}Z'^{\alpha\dot{\alpha}})^2 \nonumber \\
&-&4 (1+\varkappa^2) (P_{a\dot{a}}Y'^{a\dot{a}} +P_{\alpha\dot{\alpha}}Z'^{\alpha\dot{\alpha}})^2\Bigg), \nonumber \\
\mathcal{H}^{WZ}_4 &=& 8 i\varkappa(1+\varkappa^2)^{1\ov2} \left( Z^{3\dot{4}} Z^{4\dot{3}} ( P_{3\dot{3}} Z'^{3\dot{3}} -P_{4\dot{4}} Z'^{4\dot{4}} ) + Y^{1\dot{2}} Y^{2\dot{1}} ( P_{1\dot{1}} Y'^{1\dot{1}} -P_{2\dot{2}} Y'^{2\dot{2}} ) \right)\, . \nonumber
\eea
}

\noindent
Note that we have used the Virasoro constraint $C_1$ in order to express $x'_-$ in terms of the two index fields.
The gauge dependent terms multiplying $(2a-1)$ are invariant under SO(8) as in the underformed case.

\subsection{Tree level bosonic S-matrix}

The computation of the tree level bosonic S-matrix follows the route reviewed in \cite{Arutyunov:2009ga}, and we also use the same notations.
It is convenient to rewrite the tree-level S-matrix as a sum of two terms $\mathbb{T}=\mathbb{T}^G + \mathbb{T}^{WZ}$, coming from $\mathcal{H}^G_4$ and $\mathcal{H}^{WZ}_4$ respectevely. The reason is that $\mathbb{T}^G$ preserves the $\alg{so}(4)\oplus \alg{so}(4)$ symmetry, while $\mathbb{T}^{WZ}$ breaks it.
To write the results we always assume that $p>p'$.
Then, one finds that the action of $\mathbb{T}^G$ on the two-particle states is given by
\begin{equation}
\label{Tmatrix}
\begin{aligned}
\mathbb{T}^G \, \ket{Y_{a\dot{c}} Y_{b\dot{d}}'} &= \left[ \frac{1-2a}{2}(p \omega' - p' \omega) +\frac{1}{2} \frac{ (p-p')^2 +\dpnu^2 (\omega-\omega')^2}{p \omega' - p' \omega}  \right] \, \ket{Y_{a\dot{c}} Y_{b\dot{d}}'}  \\ 
& + \frac{p p' + \dpnu^2 \omega \omega'  }{p \omega' - p' \omega} \left( \ket{Y_{a\dot{d}} Y_{b\dot{c}}'} + \ket{Y_{b\dot{c}} Y_{a\dot{d}}'} \right), \\ \\
\mathbb{T}^G \, \ket{Z_{\alpha\dot{\gamma}} Z_{\beta\dot{\delta}}'} &=  \left[ \frac{1-2a}{2}(p \omega' - p' \omega) -\frac{1}{2} \frac{ (p-p')^2 +\dpnu^2 (\omega-\omega')^2  }{p \omega' - p' \omega} \right] \, \ket{Z_{\alpha\dot{\gamma}} Z_{\beta\dot{\delta}}'} \\
& - \frac{ p p' + \dpnu^2 \omega \omega'  }{p \omega' - p' \omega} \left( \ket{Z_{\alpha\dot{\delta}} Z_{\beta\dot{\gamma}}'} + \ket{Z_{\beta\dot{\gamma}} Z_{\alpha\dot{\delta}}'} \right), \\ \\
\mathbb{T}^G \, \ket{Y_{a\dot{b}} Z_{\alpha\dot{\beta}}'} &=  \left[ \frac{1-2a}{2}(p \omega' - p' \omega) -\frac{1}{2}\frac{\om^2-\om'^2}{p \omega' - p' \omega} \right] \, \ket{Y_{a\dot{b}} Z_{\alpha\dot{\beta}}'}, \\ \\
\mathbb{T}^G \, \ket{Z_{\alpha\dot{\beta}} Y_{a\dot{b}}'} &=  \left[ \frac{1-2a}{2}(p \omega' - p' \omega)  +\frac{1}{2}\frac{\om^2-\om'^2}{p \omega' - p' \omega} \right] \, \ket{Z_{\alpha\dot{\beta}} Y_{a\dot{b}}'},
\end{aligned}
\end{equation}
and the  action of $\mathbb{T}^{WZ}$ on the two-particle states is
\begin{equation}
\begin{aligned}
\mathbb{T}^{WZ} \, \ket{Y_{a\dot{c}} Y_{b\dot{d}}'} &
 = i \dpnu\left(\epsilon_{ab} \ket{Y_{b\dot{c}} Y_{a\dot{d}}'} 
+\epsilon_{\dot{c}\dot{d}} \ket{Y_{a\dot{d}} Y_{b\dot{c}}'}\right)
, \\
\mathbb{T}^{WZ} \, \ket{Z_{\alpha\dot{\gamma}} Z_{\beta\dot{\delta}}'} &
 = i \dpnu \left( \epsilon_{\alpha\beta} \ket{Z_{\beta\dot{\gamma}} Z_{\alpha\dot{\delta}}'} 
+ \epsilon_{\dot{\gamma}\dot{\delta}} \ket{Z_{\alpha\dot{\delta}} Z_{\beta\dot{\gamma}}'}\right)\,,
\end{aligned}
\end{equation}
where on the r.h.s. we obviously do not sum over the repeated indices. In the formulae the frequency $\omega$ is related to the momentum $p$ as 
\be\la{omega}
\omega=(1+\varkappa^2)^{1\ov2} \sqrt{1+p^2} = \sqrt{1+p^2\ov 1-\dpnu^2}\,,
\ee
and we have introduced the parameter 
\be
\dpnu = {\varkappa\ov (1+\varkappa^2)^{1\ov2}}={2\e\ov 1+\e^2}\,,
\ee
which, as one can see from the expressions above, is the natural deformation parameter. In fact, as we discuss in the next subsection, it is related in a very simple way to the parameter $q$ of the q-deformed S-matrix:  $q=e^{-\dpnu/g}$.

The S-matrix ${\mathbb S}$ computed in perturbation theory is related to the ${\mathbb T}$-matrix as 
\be\la{Tpert}
{\mathbb S}=\mI +\frac{i}{g}{\mathbb T}\, .
\ee
In the undeformed case, as a consequence of invariance of ${\mathbb S}$ with respect to two copies of the centrally extended superalgebra $\psu(2|2)$, the corresponding ${\mathbb T}$-matrix 
admits a factorization
\bea
{\mathbb T}^{P\dot{P},Q\dot{Q}}_{M\dot{M},N\dot{N}}=(-1)^{\eps_{\dot M}(\eps_{N}+\eps_{Q})}{\cal T}_{MN}^{PQ}\delta_{\dot{M}}^{\dot{P}}\delta_{\dot{N}}^{\dot{Q}}
+(-1)^{\eps_Q(\eps_{\dot{M}}+\eps_{\dot{P}})}\delta_{M}^{P}\delta_{N}^{Q} {\cal T}_{\dot{M}\dot{N}}^{\dot{P}\dot{Q}}\, .
\eea 
Here $M=(a, \alpha)$ and $\dot{M}=(\dot{a},\dot{\alpha})$, and  dotted and undotted indices are referred to two copies of $\psu(2|2)$, respectively, while 
$\eps_{M}$ and $\eps_{\dot{M}}$ describe statistics of the corresponding indices, {\it i.e.} they  are zero for bosonic (Latin) indices and equal to one for fermionic (Greek) ones.
The factor ${\mathcal T}$ can be regarded as $16\times 16$ matrix.

\smallskip

It is not difficult to see that the same type of factorization persists in the deformed case as well. Indeed, from the formulae  (\ref{Tmatrix}) we extract the following elements for the ${\mathcal T}$-matrix
\bea\la{cTmatr}
\begin{aligned}
&\cT_{ab}^{cd}=
A\,\de_a^c\de_b^d+B\,\de_a^d\de_b^c+W\, \eps_{ab}\de_a^d\de_b^c\, ,   \\ 
&\cT_{\a\b}^{\g\de}=
D\,\de_\a^\g\de_\b^\de+E\,\de_\a^\de\de_\b^\g+W\, \eps_{\a\b}\, \de_{\a}^{\delta}\de_{\b}^{\gamma}\,,  \\
&\cT_{a\b}^{c\de}= G\,\de_a^c\de_\b^\de\,,\qquad
~\cT_{\a b}^{\g d}= L\,\de_\a^\g\de_b^d\,,
\end{aligned}
\eea where the
coefficients are given by 
\bea
\begin{aligned}
\la{Tmatrcoef}
&A(p,p')= \frac{1-2a}{4}(p \omega' - p' \omega) +\frac{1}{4} \frac{ (p-p')^2 +\dpnu^2 (\omega-\omega')^2}{p \omega' - p' \omega} \,,\\
&B(p,p')=-E(p,p')= \frac{p p' + \dpnu^2 \omega \omega'  }{p \omega' - p' \omega} \,, \\
&D(p,p')=\frac{1-2a}{4} (p \omega' - p' \omega) -\frac{1}{4} \frac{ (p-p')^2 +\dpnu^2 (\omega-\omega')^2  }{p \omega' - p' \omega} \,,\\
&G(p,p')=-L(p',p)=\frac{1-2a}{4}(p \omega' - p' \omega) -\frac{1}{4} \frac{\om^2-\om'^2}{p \omega' - p' \omega} \,,\\
&W(p,p')= i\dpnu   \, .
\end{aligned}
\eea
Here $W$ corresponds to the contribution of the Wess-Zumino term and it does not actually depend on the particle momenta.
All the four remaining coefficients $\cT_{ab}^{\g\de},\cT_{\a\b}^{cd},\cT_{a\b}^{\g d},\cT_{\a b}^{\g d}$ vanish in the bosonic case but will be switched on once fermions are
taken into account. The matrix ${\mathcal T}$ is recovered from its matrix elements as follows
\bea
\nonumber
{\mathcal T}={\cal T}_{MN}^{PQ}\, E_P^M\otimes E_Q^N=\cT_{ab}^{cd}\, E_c^a\otimes E_d^b+\cT_{\a\b}^{\g\de}\, E_\g^\a\otimes E_\delta^\b+
\cT_{a\b}^{c\de}\, E_c^a\otimes E_\de^\b+\cT_{\a b}^{\g d}\, E_\g^\a\otimes E_d^b\, ,
\eea
where $E_M^N$ are the standard matrix unities. For the reader convenience we present ${\mathcal T}$ as an explicit $16\times 16$ matrix\footnote{See appendix 8.5 of 
\cite{Arutyunov:2006yd}
for the corresponding matrix in the undeformed case.}

{\scriptsize
\begin{eqnarray}
{\mathcal T}\equiv \left( \begin{array}{ccccccccccccccccccc}
{\cal A}_1&0&0&0&|&0&0&0&0&|&0&0&0&0&|&0&0&0&0\\
0&{\cal A}_2&0&0&|&{\cal A}_4&0&0&0&|&0&0&0&0&|&0&0&0 &0\\
0&0&{\cal A}_3&0&|&0&0&0&0&|&0&0&0&0&|&0&0&0&0\\
0&0&0&{\cal A}_3&|&0&0&0&0&|&0&0&0&0&|&0&0&0&0\\
-&-&-&-&-&-&-&-&-&-&-&-&-&-&-&-&-&-&-\\
0&{\cal A}_5&0&0&|& {\cal A}_2&0&0&0&|&0&0&0&0&|&0&0& 0 &0\\
0&0&0&0&|&0&{\cal A}_1&0&0&|&0&0&0&0&|&0&0&0&0\\
0&0&0&0&|&0&0&{\cal A}_3&0&|&0&0&0&0&|&0&0&0&0\\
0&0&0&0&|&0&0&0&{\cal A}_3&|&0&0&0&0&|&0&0&0&0\\
-&-&-&-&-&-&-&-&-&-&-&-&-&-&-&-&-&-&-\\
0&0&0&0&|&0&0&0&0&|&{\cal A}_8&0&0&0&|&0&0&0&0\\
0&0&0&0&|&0&0&0&0&|&0&{\cal A}_8&0&0&|&0&0&0&0\\
0&0&0&0&|&0&0&0&0&|&0&0& {\cal A}_6 &0&|&0&0&0&0\\
0&0&0&0&|& 0&0&0&0&|&0&0&0&{\cal A}_7&|&0&0&{\cal A}_9&0\\
-&-&-&-&-&-&-&-&-&-&-&-&-&-&-&-&-&-&-\\
0&0&0&0&|&0&0&0&0&|&0&0&0&0&|&{\cal A}_8&0&0&0\\
0&0&0&0&|&0&0&0&0&|&0&0&0&0&|&0&{\cal A}_8&0&0\\
0&0&0&0&|&0&0&0&0&|&0&0&0&{\cal A}_{10}&|&0&0&{\cal A}_7&0\\
0&0&0&0&|&0&0&0&0&|&0&0&0&0&|&0&0&0&{\cal A}_6\\
\end{array} \right) \, .\nonumber
\end{eqnarray}
 }
Here the non-trivial matrix elements of ${\mathcal T}$ are given by
 \bea
 &&{\cal A}_1=A+B\, ,\quad
 {\cal A}_2=A\, ,\quad
 {\cal A}_4=B-W\, ,\quad
 {\cal A}_5=B+W\,, \quad{\cal A}_6=D+E\, ,
 \\\nonumber
 &&{\cal A}_6=D+E\, ,\quad
 {\cal A}_7=D\, , \quad
 {\cal A}_8=L\, ,\quad
 {\cal A}_9=E-W=-{\cal A}_5 \, ,\quad
 {\cal A}_{10}=E+W=-{\cal A}_4\, .
 \eea

 We conclude this section by pointing out that the found matrix ${\mathcal T}$ satisfies the classical Yang-Baxter equation
\bea
[\cT_{12}(p_1,p_2),\cT_{13}(p_1,p_3)+\cT_{23}(p_2,p_3)]+[\cT_{13}(p_1,p_3),\cT_{23}(p_2,p_3)]=0\, 
\eea  
for any value of the deformation parameter $\dpnu$.

\subsection{Comparison with the q-deformed S-matrix}

In this subsection we show that the perturbative bosonic world-sheet S-matrix coincides with the first nontrivial term in the large $g$ expansion of the q-deformed \ads S-matrix, in other words with the corresponding classical $r$-matrix\footnote{The difference with the expansion performed in \cite{Beisert:2010kk} is that we include the dressing factor in the definition of the S-matrix.}.

 Let us recall that up to an overall factor the q-deformed \ads S-matrix is given by a tensor product of two copies of the $\psu(2|2)_q$-invariant S-matrix \cite{Beisert:2008tw} which is reviewed in appendix \ref{app:matrixSmatrix}. 
Including the overall factor $S_{\su(2)}$ which is the scattering matrix in the $\su(2)$ sector, the complete S-matrix can be written in the form \cite{Ben}
\bea
&&\hspace{-1.5cm}
{\mathbf S}=S_{\su(2)} S\, \hat{\otimes} \, S\, , ~~~S_{\su(2)}=\frac{e^{i a(p_2{\cal E}_1-p_1{\cal E}_2)}}{\sigma_{12}^2}\frac{x_1^++\xi}{x_1^-+\xi}\frac{x_2^-+\xi}{x_2^++\xi}\cdot
\frac{x_1^--x_2^+}{x_1^+-x_2^-}\frac{1-\frac{1}{x_1^-x_2^+}}{1-\frac{1}{x_1^+x_2^-}}\, ,
\eea
where $S$ is the $\psu(2|2)_q$-invariant S-matrix \eqref{Sqmat},   $\hat{\otimes} $ stands for the graded tensor product, $a$ is the parameter of the light-cone gauge \eqref{lcg}, $\sigma$ is the dressing factor, and ${\cal E}$ is the q-deformed dispersion relation \eqref{qdisp} whose large $g$ expansion starts with $\om$.  The dressing factor can be found by solving the corresponding crossing equation,  and  it is given by \cite{Ben}
\begin{equation}\label{eq:def-theta}
\sigma_{12}=e^{i\t_{12}}\,,\quad \theta_{12} = \chi(x^+_1,x^+_2) + \chi(x^-_1,x^-_2)- \chi(x^+_1,x^-_2) - \chi(x^-_1,x^+_2),
\end{equation}
where
\begin{equation}\la{chi12}
\chi(x_1,x_2) = i \oint \frac{dz}{2 \pi i} \frac{1}{z-x_1} \oint \frac{dz'}{2 \pi i} \frac{1}{z'-x_2} \log\frac{\Gamma_{q^2}(1+\frac{i g}{2} (u(z)-u(z')))}{\Gamma_{q^2}(1-\frac{i g}{2} (u(z)-u(z')))}.
\end{equation}
Here $\Gamma_{q}(x)$ is the q-deformed Gamma function which for complex $q$
 admits an integral representation  \eqref{lnGovG} \cite{Ben}.

To develop the large $g$ expansion of the q-deformed \ads S-matrix, one  has to assume that $q=e^{-\qdp/g}$ where $\qdp$ is a deformation parameter which is kept fixed in the limit $g\to\infty$, and should be related to $\dpnu$. Then, due to the factorisation of the perturbative bosonic world-sheet S-matrix and the q-deformed \ads S-matrix, it is sufficient to compare the ${\cal T}$-matrix \eqref{cTmatr} with the ${\mathbf T}$-matrix appearing in the expansion of the ``square root'' of $\mathbf S$
\be\la{qTpert}
S_{\su(2)}^{1/2}\, \mI_g\, S=\mI +\frac{i}{g}{\mathbf T}\, ,
\ee
where $\mI_g$ is the graded identity which is introduced so that the expansion starts with $\mI$. 
The only term which is not straightforward to expand is the $S_{\su(2)}$ scalar factor because it contains the dressing phase $\t_{12}$.  
The scalar factor obviously can contribute only to the part of the ${\mathbf T}$-matrix proportional to the identity matrix. Since in the expansion of the $\psu(2|2)_q$-invariant S-matrix \eqref{Sqmat}, $\mI_gS=\mI+{i\ov g}r$,  the element $r_{11}^{11}$ is equal to 0 (because $a_1=1$) it is convenient to subtract ${\cal T}_{11}^{11}\mI = {\cal A}_1\mI$ from the ${\cal T}$-matrix and compare the resulting matrix with the classical $r$-matrix. One should obviously remove the off-diagonal terms from the classical $r$-matrix which appear due to the presence of fermions in the full superstring action \eqref{defLag}. 
With this done, one finds that they are equal to each other provided $\qdp = \dpnu$, and therefore $q=e^{-\dpnu/g}$ is real. Thus, to show that ${\mathbf T}={\cal T}$ one should demonstrate that 
$ {\cal A}_1$ is equal to the $1/g$ term in the expansion of $S_{\su(2)}^{1/2}$. 
To this end one should find the large $g$ expansion of the dressing phase $\t_{12}$ which is done by first expanding the ratio of $\Gamma_{q^2}$-functions in \eqref{chi12} with $u(z)$ and $u(z')$ being kept fixed. This is done in appendix  \ref{app:qGamma}, see \eqref{qGamma1}.
Next, one combines it with the expansion of the $\frac{1}{z-x_1^\pm} \frac{1}{z'-x_2^\pm} $ terms which appear in the integrand of \eqref{eq:def-theta}. As a result one finds that the dressing phase is of order $1/g$ just as it was in the undeformed case \cite{AFS}.  We have not tried to compute the resulting double integrals analytically but we have checked numerically that the element 
 ${\cal A}_1$ is indeed equal to the $1/g$ term in the expansion of $S_{\su(2)}^{1/2}$ if the deformation parameter $\dpnu$ satisfies $\nu<1/\sqrt 2$. At $\nu=1/\sqrt 2$ the integral representation for the dressing factor breaks down but it is unclear to us if it is a signal of a genuine problem with the q-deformed S-matrix.   In fact it is not difficult to extract from  ${\cal A}_1$ the leading term in the large $g$ expansion of the dressing phase which appears to be very simple
 \be
\t_{12}= \frac{\dpnu^2 \left(\omega _1-\omega _2\right)+p_2^2 \left(\omega _1-1\right)-p_1^2 \left(\omega _2-1\right)}{2g
   \left(p_1+p_2\right)} +\cdots\,.
\ee
It would be curious to derive this expression from the double integral representation. Note that doing this double integral one also could get the full AFS order of the phase.

\section{Conclusions}
In this work we successfully matched in the large tension limit the tree-level bosonic S-matrix arising from the sigma-model on the deformed \ads  space with the 
q-deformed S-matrix obtained from symmetries. There are many other important issues to be addressed. 

\smallskip

 We identified NSNS background fields in the string frame.  More studies are needed however to  extract RR fields since the latter couple directly to fermionic degrees of freedom. 
 Rather intricate field redefinitions should be performed to bring the deformed action to the standard one for Type IIB superstring in an arbitrary supergravity background,
  thus allowing the identification of the full bosonic background. It might be easier in fact just to use the NSNS background fields and the type IIB supergravity equations of motion to find the full supergravity background \cite{ABF}.
  
 \smallskip

Next, the matching of S-matrices, successful at  tree level, can be further extended  by computing admittedly more complicated loop corrections to the tree-level scattering matrix 
 of the light-cone sigma-model; this also requires taking fermions into account. It is natural to expect that the deformation parameter $\dpnu$  undergoes a non-trivial renormalization to fit the parameter $q$ entering the exact, {\it i.e.}~all-loop, 
 q-deformed S-matrix.

\smallskip

We also showed that in the large tension limit the conjectured dispersion relation \eqref{qdisp} turns into the perturbative one  \eqref{omega}. It would be interesting to find an $\eta$-deformed giant magnon solution \cite{HM} which would provide additional evidence in favour of \eqref{qdisp}. In the case of the finite angular momentum the corresponding solution would also provide important information about the structure of the finite size corrections \cite{AFZ} in the $\eta$-deformed theory.

\smallskip

It is also interesting to find explicit spinning string solutions of  the $\dpo$-deformed bosonic sigma-model. Due to the singularity of the $\dpo$-deformed AdS  a particularly interesting solution to analyse would be the GKP string and its generalisation \cite{GKP,FT02}. Then, in the case of $\AdS$, substituting the spinning Ansatz in the sigma-model
 equations of motion  leads to the emergence \cite{Arutyunov:2003uj} of the Neumann model, a famous finite-dimensional integrable system. One may hope that studies of the $\dpo$-deformed sigma-model in this context may reveal new integrable finite-dimensional systems which can be described as deformations of  the Neumann model.
 Furthermore, known finite-gap integration techniques can be applied to obtain a wider class of solutions that generalize the solutions of the Neumann system. Normally they are 
 described by a certain algebraic curve  which is supposed to emerge from the Bethe Ansatz based on the exact q-deformed S-matrix in the semi-classical limit. This would serve as another 
 non-trivial check that the two models, one based on the explicitly known deformed action  and the other based on the exact quantum group symmetry, have a good chance to describe the same physics.

\smallskip

 One can also adopt the logic of the undeformed case construction and use the exact q-deformed S-matrix to engineer the mirror TBA equations for real $q$; a solution of this problem is under way
 \cite{Arutyunov:2014ota}.

\smallskip

With the knowledge of a complete supergravity background and its symmetries for the deformed case, 
one can approach perhaps the most interesting question about the dual gauge theory. 
Since the deformation affects the isometries of the AdS space, the theory will be neither conformal nor Lorentz invariant. Since there is a $B$-field on the string theory side,
one may expect that this theory is a non-commutative deformation of ${\cal N}=4$ super Yang-Mills in the sense of the Moyal star product
with a hidden quantum group symmetry which would include the two copies of the
$\psu_q(2|2)$ algebra.
It would be fascinating to construct such a theory explicitly.

\bigskip

\section*{Acknowledgments}


\noindent
We thank Marius de Leeuw, Stijn van Tongeren and Benoit Vicedo for useful discussions.
G.A. and R.B. acknowledge support by the Netherlands Organization for Scientific Research (NWO) under the VICI grant 680-47-602.
The work by G.A. and R.B. is also a part of the ERC Advanced grant research programme No. 246974,  {\it ``Supersymmetry: a
 window to non-perturbative physics"} and of the D-ITP consortium, a program of the NWO that is funded by the Dutch Ministry of Education, Culture and Science (OCW).
S.F. is supported by a DFG grant in the framework of the SFB 647 ``Raum - Zeit - Materie. Analytische und Geometrische Strukturen''
and by the Science Foundation Ireland under Grant 09/RFP/PHY2142.

\appendix

\section{The inverse operator and bosonic Lagrangian}\label{IRgd}

To find the bosonic part of the deformed Lagrangian one needs to choose a coset representative $\ag$, and invert the operator  $1-\e R_\ag\circ d$. 
We find useful the
following  parametrisation  of a bosonic coset element 
\be \label{basiccoset}
\ag_{\alg{b}}=\small{\left(
\begin{array}{cc}
 \ag_{\alg{a}}
& 0
\\
0 &
 \ag_{\alg{s}}
\end{array}
\right)}\, ,\quad \ag_{\alg{a}}=\Lambda(\psi_k)\, \Xi(\z)\check\ag_{\r}(\r)\, ,\quad \ag_{\alg{s}}=\Lambda(\p_k)\, \Xi(\xi)\check\ag_{\rm r}(r)\, . \ee 
Here the matrix functions $\Lambda$, $\Xi$ and $\check\ag$ are defined as 
\be \label{Lambda}
\Lambda(\vp_k)=\exp(\sum_{k=1}^3
\frac{i}{2}\vp_k h_k )\,,\quad  \Xi(\vp)=\left(
\begin{array}{cccc}
 \cos\frac{\vp}{2} & \sin\frac{\vp}{2} & 0 & 0 \\
 -\sin\frac{\vp}{2} & \cos\frac{\vp}{2} & 0 & 0 \\
 0 & 0 & \cos\frac{\vp}{2} & -\sin\frac{\vp}{2} \\
 0 & 0 & \sin\frac{\vp}{2} & \cos\frac{\vp}{2} \\
\end{array}
\right)\,, 
\ee
\be \label{checkgrho}
 \check\ag_{\rho}(\r) =
\left(
\begin{array}{cccc}
 \r_+& 0 & 0 &\r_- \\
 0 & \r_+& -\r_-& 0 \\
 0 & -\r_- & \r_+& 0 \\
\r_-& 0 & 0 & \r_+\\
\end{array}
\right) \,,\quad \r_\pm= {\sqrt{\sqrt{\rho ^2+1}\pm1} \ov\sqrt 2}\,,\ee
\be \label{checkgr}
 \check\ag_{r}(r) =
\left(
\begin{array}{cccc}
 r_+& 0 & 0 &i\, r_- \\
 0 & r_+& -i \,r_-& 0 \\
 0 & -i\,r_- & r_+& 0 \\
i\,r_-& 0 & 0 & r_+\\
\end{array}
\right) \,,\quad r_\pm= {\sqrt{1\pm\sqrt{1-r ^2}} \ov\sqrt 2}\,,\ee
where the diagonal matrices $h_i$ are given by
\be
h_1={\rm diag}(-1,1,-1,1) \,,\quad 
h_2={\rm diag}(-1,1,1,-1) \,,\quad
h_3={\rm diag}(1,1,-1,-1) \,.
\ee
In the undeformed case the ${\rm AdS}_5$ coordinates $t\equiv\psi_3\,,\,\psi_1\,,\,\psi_2\,,\, \z\,,\, \r$ are related to the embedding 
coordinates $Z^A,$ $A = 0,1,\ldots,5$ obeying the constraint $\eta^{AB}Z_AZ_B=-1$ where $\eta^{AB}=(-1,1,1,1,1,-1)$ as follows
\bea
Z_1+iZ_2 = \r\cos\z\,e^{i\psi_1}\,,\quad Z_3+iZ_4 = \r\sin\z\,e^{i\psi_2}\,,\quad Z_0+iZ_5 = \sqrt{1+\r^2}\,e^{it}\,,
\eea
while the ${\rm S}_5$ coordinates $\phi\equiv\p_3\,,\,\p_1\,,\,\p_2\,,\, \xi\,,\, r$ are related to the embedding 
coordinates $Y^A,$ $A = 1,\ldots,6$ obeying $Y_A^2=1$  as
\bea
Y_1+iY_2 = r\cos\xi\,e^{i\p_1}\,,\quad Y_3+iY_4 = r\sin\xi\,e^{i\p_2}\,,\quad Y_5+iY_6 = \sqrt{1-r^2}\,e^{i\p}\,.
\eea
An important property of the coset representative \eqref{basiccoset} is that the $R_\ag$ operator is independent of the angles $\psi_k$ and $\p_k$:
\be
R_{\ag_{\alg{b}}}(M) = R_{\check\ag}(M) \,,\quad \check\ag=\small{\left(
\begin{array}{cc}
 \check\ag_{\alg{a}}
& 0
\\
0 &
 \check\ag_{\alg{s}}
\end{array}
\right)}\, ,\quad \check\ag_{\alg{a}}=\Xi(\z)\check\ag_{\r}(\r)\, ,\quad \check\ag_{\alg{s}}=\Xi(\xi)\check\ag_{\rm r}(r)\, . \ee 
To compute the Lagrangian one needs to know the action of the operator $1/(1-\e R_\ag\circ d)$ on the projection $M^{(2)}$ and $M_{\rm odd}$ of an $\su(2|2)$ element $M$. 
This action on odd elements appears to be $\check\ag$-independent
\be
{1\ov 1-\e R_{\check\ag}\circ d}(M_{\rm odd}) ={\mI + \eta R\circ d\ov 1 - \eta^2}(M_{\rm odd}) \,.
\ee
This action on $M^{(2)}$ factorizes into a sum of actions on $M_{\alg{a}}$ and $M_{\alg{s}}$ where $M_{\alg{a}}$ is the upper left $4\times4$ block of $M^{(2)}$, and $M_{\alg{s}}$  is the lower right $4\times4$ block of $M^{(2)}$. One can check that the inverse operator is given by
\be
{1\ov 1-\e R_{\check\ag}\circ d}(M_{\alg{a}}) =\Big(\mI+
{\e^3f_{31}^{\alg{a}}+\e^4f_{42}^{\alg{a}}+\e^5h_{53}^{\alg{a}}\ov (1-c_{\alg{a}}\e^2)(1-d_{\alg{a}}\e^2)} +
{\e R_{\check\ag}\circ d + \e^2 R_{\check\ag}\circ d\circ R_{\check\ag}\circ d\ov 1-c_{\alg{a}}\e^2}\Big)\big(M_{\alg{a}}\big)\,,
\ee
\be
{1\ov 1-\e R_{\check\ag}\circ d}(M_{\alg{s}}) =\Big(\mI+
{\e^3f_{31}^{\alg{s}}+\e^4f_{42}^{\alg{s}}+\e^5h_{53}^{\alg{s}}\ov (1-c_{\alg{s}}\e^2)(1-d_{\alg{s}}\e^2)} +
{\e R_{\check\ag}\circ d + \e^2 R_{\check\ag}\circ d\circ R_{\check\ag}\circ d\ov 1-c_{\alg{s}}\e^2}\Big)\big(M_{\alg{s}}\big)\,.
\ee
Here 
\be
c_{\alg{a}}= \frac{4\rho^2}{\left(1-\eta ^2\right)^2}\,,\quad d_{\alg{a}}=-\frac{4\rho^4 \sin^2\z }{\left(1-\eta^2\right)^2} \,,\quad c_{\alg{s}}= -\frac{4r^2}{\left(1-\eta ^2\right)^2}\,,\quad d_{\alg{s}}=-\frac{4r^4 \sin^2\xi }{\left(1-\eta^2\right)^2} \,,
\ee
\be
f_{k,k-2}^{\alg{a}}(M_{\alg{a}}) =\Big(\big(R_{\check\ag}\circ d\big)^k - c_{\alg{a}}\big(R_{\check\ag}\circ d\big)^{k-2}\Big)(M_{\alg{a}}) \,,
\ee
\be
f_{k,k-2}^{\alg{s}}(M_{\alg{s}}) =\Big(\big(R_{\check\ag}\circ d\big)^k - c_{\alg{s}}\big(R_{\check\ag}\circ d\big)^{k-2}\Big)(M_{\alg{s}}) \,,
\ee
$d_{\alg{a}}$ and $d_{\alg{s}}$ appear in the identities
\be
f_{k+2,k}^{\alg{a}} = d_{\alg{a}} f_{k,k-2}^{\alg{a}} \,,\quad f_{k+2,k}^{\alg{s}} = d_{\alg{s}} f_{k,k-2}^{\alg{s}} \,,\quad k = 4,5,\ldots\,,
\ee
and $h_{53}^{\alg{a}}$ and $h_{53}^{\alg{s}}$ appear in 
\be
h_{53}^{\alg{a}}=f_{53}^{\alg{a}} - d_{\alg{a}} f_{31}^{\alg{a}} \,,\quad h_{53}^{\alg{s}}=f_{53}^{\alg{s}} - d_{\alg{s}} f_{31}^{\alg{s}} \,.
\ee

The bosonic Lagrangian can then be easily computed and is given by (\ref{Lfull}-\ref{LsWZ}). To find the quartic Lagrangian used for 
computing the bosonic part of the four-particle world-sheet scattering matrix, we first expand the Lagrangian \eqref{Lfull} up to quartic order in $\r$, $r$ and their derivatives  
\bea
  \begin{aligned}
\L_{\alg{a}} &=-{g\ov2}(1+\varkappa^2)^{1\ov2}\Big( \g^{\a\b}\Big[-\pa_\a t\pa_\b t (1+(1+\varkappa ^2) \rho ^2 (1+\varkappa ^2 \rho
   ^2))+
\pa_\a \r\pa_\b \rho  (1+(\varkappa ^2-1) \rho ^2)     \\
&+
\pa_\a \psi_1\pa_\b\psi_1\rho ^2 \cos
   ^2\z+\pa_\a \psi_2\pa_\b\psi_2
  \rho ^2 \sin ^2\z+\pa_\a \z\pa_\b\z \rho ^2\Big]- \varkappa \eps^{\a\b} \rho ^4 \sin 2 \zeta\pa_\a\psi_1\pa_\b\zeta\Big)\,, \\
  \\
\L_{\alg{s}} &=-{g\ov2}(1+\varkappa^2)^{1\ov2}\Big(  \g^{\a\b}\Big[\pa_\a \p\pa_\b \p
  (1-(1+\varkappa ^2) r^2 (1-\varkappa ^2 r^2))    +\pa_\a r\pa_\b r  (1+(1-\varkappa^2)r^2)  \\
   &+
   \pa_\a \p_1\pa_\b \p_1  r^2 \cos ^2\xi +\pa_\a \p_2\pa_\b \p_2  r^2 \sin^2\xi +\pa_\a \xi\pa_\b \xi  r^2 \Big]+ \varkappa \eps^{\a\b} r^4 \sin 2 \xi \pa_\a\p_1\pa_\b\xi\Big)\, .
\end{aligned}
\eea
Further, we make a shift 
\bea
\label{shift}
\rho\to \rho-\frac{\varkappa^2}{4}\rho^3\, , ~~~~~r\to r+\frac{\varkappa^2}{4}r^3\,\eea
so that the quartic action acquires the form 
 \bea\nonumber
\L_{\alg{a}} &=&-{g\ov2}(1+\varkappa^2)^{1\ov2}\, \g^{\a\b}\Big[-\pa_\a t\pa_\b t \Big(1+(1+\varkappa ^2)\rho^2 +\tfrac{1}{2}\varkappa^2 (1+\varkappa ^2) \rho^4\Big)+\\
&&\hspace{-1cm}+\pa_\a \r\pa_\b \rho  \Big(1- \rho ^2-\tfrac{\varkappa^2}{2}\rho^4\Big) +\Big(\rho ^2-\tfrac{\varkappa^2}{2}\rho^4\Big) \Big(\pa_\a \psi_1\pa_\b\psi_1\cos
   ^2\z+\pa_\a \psi_2\pa_\b\psi_2
  \sin ^2\z+\pa_\a \z\pa_\b\z \Big)\Big] \nonumber  \\
&&\hspace{-1cm}+{g\ov2} \varkappa (1+\varkappa^2)^{1\ov2}\eps^{\a\b} \rho ^4 \sin 2 \zeta\pa_\a\psi_1\pa_\b\zeta\,, 
  \eea
\bea
\L_{\alg{s}} &=&-{g\ov2}(1+\varkappa^2)^{1\ov2}\, \g^{\a\b}\Big[\pa_\a \p\pa_\b \p
 \Big(1-(1+\varkappa ^2) r^2+\tfrac{1}{2}\varkappa^2(1+\varkappa^2)r^4\Big)    + \nonumber \\
&&\hspace{-1cm}+\pa_\a r\pa_\b r  \Big(1+r^2+\tfrac{\varkappa^2}{2}r^4\Big) +  \Big(r^2+ \tfrac{\varkappa^2}{2}r^4\Big)\Big(\pa_\a \p_1\pa_\b \p_1   \cos ^2\xi +\pa_\a \p_2\pa_\b \p_2  \sin^2\xi +\pa_\a \xi\pa_\b \xi  \Big)\Big] \nonumber \\
&&\hspace{-1cm}-{g\ov2} \varkappa(1+\varkappa^2)^{1\ov2} \eps^{\a\b} r^4 \sin 2 \xi \pa_\a\p_1\pa_\b\xi\, .\eea
Changing the spherical coordinates to $(z_i,y_i)$, see \eqref{flatcoord}, 
and expanding the resulting action up to the quartic order in $z$ and $y$ fields we get the quartic Lagrangian \eqref{Lquart}. Notice that the shifts of $\r$ and $r$ in \eqref{shift} were chosen so that the deformed metric expanded up to quadratic order in the fields would be diagonal.

\medskip

It is also possible to choose a coset representative precisely in the same way as is done in the undeformed case, see \cite{Arutyunov:2009ga} for details. 
Accordingly, for the metric pieces we obtain 
\bea
\label{adsL}
\L_{\alg{a}}^{G}&=&-\frac{g}{2}(1+\varkappa^2)^{1\ov2}\gamma^{\a\b}\Big[-G_{tt}\pa_{\a}t\pa_{\beta}t+G_{zz}\pa_{\a}z_i\pa_{\beta}z_i+G_{\alg{a}}^{(1)}z_i\pa_{\a}z_iz_j\pa_{\b}z_j+\nonumber \\
&&~~~~~~~~~~~~~~~~~~~~~~~~~~ +G_{\alg{a}}^{(2)}(z_3\pa_{\a}z_4-z_4\pa_{\a}z_3)(z_3\pa_{\b}z_4-z_4\pa_{\b}z_3)
\Big]\, ,  \\
\label{spL}
\L_{\alg{s}}^{G}&=&-\frac{g}{2}(1+\varkappa^2)^{1\ov2}\gamma^{\a\b}\Big[G_{\phi\phi}\pa_{\a}\phi\pa_{\beta}\phi+G_{yy}\pa_{\a}y_i\pa_{\beta}y_i+G_{\alg{s}}^{(1)}y_i\pa_{\a}y_iy_j\pa_{\b}y_j+\nonumber \\
&&~~~~~~~~~~~~~~~~~~~~~~~~~~ +G_{\alg{s}}^{(2)}(y_3\pa_{\a}y_4-y_4\pa_{\a}y_3)(y_3\pa_{\b}y_4-y_4\pa_{\b}y_3)
\Big]\, .
\eea
Here the coordinates $z_i$, $i=1,\ldots,4$, and $t$ parametrize the deformed AdS space, while the coordinates $y_i$, $i=1,\ldots,4$, and the angle $\phi$ parametrize the deformed five-sphere.
The components of the deformed AdS metric in (\ref{adsL}) are\footnote{Note that the coordinates $y_i$ and $z_i$ are different from the ones appearing in the quartic Lagrangian \eqref{Lquart} because the nondiagonal components of the deformed metric do not vanish. }
\bea
\begin{aligned}
\label{Gads}
\hspace{-0.5cm}
G_{tt}&=\frac{(1+z^2/4)^2}{(1-z^2/4)^2-\varkappa^2 z^2}\, , ~~~~~~~~~~~~~~~~~~~
G_{zz}=\frac{(1-z^2/4)^2}{(1-z^2/4)^4+\varkappa^2 z^2(z_3^2+z_4^2)} \, ,
 \\
G_{\alg{a}}^{(1)}&=\varkappa^2G_{tt}G_{zz}\frac{z_3^2+z_4^2+(1-z^2/4)^2}{(1-z^2/4)^2(1+z^2/4)^2}\, , ~~~~~~
G_{\alg{a}}^{(2)}=\varkappa^2 G_{zz}\frac{z^2}{(1-z^2/4)^4}
\, .
\end{aligned}
\eea
For the sphere part the corresponding expressions read
\bea
\begin{aligned}
\label{Gsphere}
\hspace{-0.5cm}
G_{\phi\phi}&=\frac{(1-y^2/4)^2}{(1+y^2/4)^2+\varkappa^2 y^2}\, , ~~~~~~~~~~~~~~~~~~~
G_{yy}=\frac{(1+y^2/4)^2}{(1+y^2/4)^4+\varkappa^2 y^2(y_3^2+y_4^2)} \, ,
 \\
G_{\alg{s}}^{(1)}&=\varkappa^2G_{\phi\phi}G_{yy}\frac{y_3^2+y_4^2-(1+y^2/4)^2}{(1-y^2/4)^2(1+y^2/4)^2}\, , ~~~~~~
G_{\alg{s}}^{(2)}=\varkappa^2 G_{yy}\frac{y^2}{(1+y^2/4)^4}
\, .
\end{aligned}
\eea
Obviously, in the limit $\varkappa\to 0$ the components $G_{\alg{a}}^{(i)}$ and $G_{\alg{s}}^{(i)}$ vanish, and one obtains the metric of the ${\rm AdS}_5\times {\rm S}^5$, {\it c.f.} fomulae (1.145) and (1.146)
in  \cite{Arutyunov:2009ga}.
Finally, for the Wess-Zumino terms the results (up to total derivative terms which do not contribute to the action) are
\bea
\begin{aligned}
\label{WZ}
\mathscr{L}_{\alg{a}}^{WZ}&=2g\varkappa(1+\varkappa^2)^{1\ov2}\, \eps^{\a\b}\frac{(z_3^2+z_4^2)\pa_{\a}z_1\pa_{\b}z_2}{(1-z^2/4)^4+\varkappa^2 z^2(z_3^2+z_4^2)}\, \\
\mathscr{L}_{\alg{s}}^{WZ}&=-2g\varkappa(1+\varkappa^2)^{1\ov2}\, \eps^{\a\b}\frac{(y_3^2+y_4^2)\pa_{\a}y_1\pa_{\b}y_2}{(1+y^2/4)^4+\varkappa^2 y^2(y_3^2+y_4^2)}\, .
\end{aligned}
\eea 

To complete our discussion of the bosonic Lagrangian of the deformed theory, let us note that in the undeformed case the action is invariant with respect to two copies of ${\rm SO}(4)$ acting linearly on $z_i$ and $y_i$ respectively. 
As is seen from the expressions above, this symmetry is broken down to four copies of ${\rm SO}(2)\sim {\rm U}(1)$. Thus, together with the two ${\rm U}(1)$ isometries acting on 
$t$ and $\phi$ the deformed action is invariant under ${\rm U}(1)^3\times {\rm U}(1)^{3}$.

\section{The $\psu(2|2)_q$-invariant S-matrix}
\label{app:matrixSmatrix}

The S-matrix compatible with $\psu(2|2)_q$ symmetry \cite{Beisert:2008tw} has been studied in detail in the recent papers  \cite{Ben,Arutyunov:2012zt,Arutyunov:2012ai,Hoare:2013ysa}.
To make the present paper self-contained, in this appendix we recall its explicit form following the same notation as in \cite{Arutyunov:2012zt}.

Let $E_{ij}\equiv E_i^j$ stand for the standard matrix unities, $i,j=1,\ldots, 4$. We introduce the following definition
\begin{equation}
E_{kilj}=(-1)^{\epsilon(l)\epsilon(k)}E_{ki}\otimes E_{lj}\, ,
\end{equation}
where $\epsilon(i)$ denotes the parity of the index, equal to $0$ for $i=1,2$ (bosons) and to $1$ for $i=3,4$ (fermions). The matrices $E_{kilj}$
are convenient to write down invariants with respect to the action of copies of $\su_q(2)\subset \psu_q(2|2)$. If we introduce
\bea
\Lambda_1&=&E_{1111}+\frac{q}{2}E_{1122}+\frac{1}{2}(2-q^2)E_{1221}+\frac{1}{2}E_{2112}+\frac{q}{2}E_{2211}+E_{2222}\, ,\nonumber\\
\Lambda_2&=&\frac{1}{2}E_{1122}-\frac{q}{2}E_{1221}-\frac{1}{2q}E_{2112}+\frac{1}{2}E_{2211}\, , \nonumber \\
\Lambda_3&=&E_{3333}+\frac{q}{2}E_{3344}+\frac{1}{2}(2-q^2)E_{3443}+\frac{1}{2}E_{4334}+\frac{q}{2}E_{4433}+E_{4444} \, , \nonumber\\
\Lambda_4&=&\frac{1}{2}E_{3344}-\frac{q}{2}E_{3443}-\frac{1}{2q}E_{4334}+\frac{1}{2}E_{4433}\, , \nonumber\\
\Lambda_5&=&E_{1133}+E_{1144}+E_{2233}+E_{2244}\, ,\\
\Lambda_6&=&E_{3311}+E_{3322}+E_{4411}+E_{4422}\, , \nonumber\\
\Lambda_7&=&E_{1324}-qE_{1423}-\frac{1}{q}E_{2314}+E_{2413}\, , \nonumber\\
\Lambda_8&=&E_{3142}-qE_{3214}-\frac{1}{q}E_{4132}+E_{4231}\, , \nonumber\\
\Lambda_9&=&E_{1331}+E_{1441}+E_{2332}+E_{2442}\, , \nonumber\\
\Lambda_{10}&=&E_{3113}+E_{3223}+E_{4114}+E_{4224}\, , \nonumber
\eea
the S-matrix of the q-deformed model is given by
\be\la{Sqmat}
S_{12}(p_1,p_2)=\sum_{k=1}^{10}a_k(p_1,p_2)\Lambda_k\, ,
\ee
where the coefficients are
\bea
a_1&=&1\, ,  \nonumber \\
a_2&=&-q+\frac{2}{q}\frac{x^-_1(1-x^-_2x^+_1)(x^+_1-x^+_2)}{x^+_1(1-x^-_1x^-_2)(x^-_1-x^+_2)}\nonumber \\
a_3&=&\frac{U_2V_2}{U_1V_1}\frac{x^+_1-x^-_2}{x^-_1-x^+_2}\nonumber \\
a_4&=&-q\frac{U_2V_2}{U_1V_1}\frac{x^+_1-x^-_2}{x^-_1-x^+_2}+\frac{2}{q}\frac{U_2V_2}{U_1V_1}\frac{x^-_2(x^+_1-x^+_2)(1-x^-_1x^+_2)}{x^+_2(x^-_1-x^+_2)(1-x^-_1x^-_2)}\nonumber \\
a_5&=&\frac{x^+_1-x^+_2}{\sqrt{q}\, U_1V_1(x^-_1-x^+_2)} \nonumber \\
a_6&=&\frac{\sqrt{q}\, U_2V_2(x^-_1-x^-_2)}{x^-_1-x^+_2}  \\
a_7&=&\frac{ig}{2}\frac{(x^+_1-x^-_1)(x^+_1-x^+_2)(x^+_2-x^-_2)}{\sqrt{q}\, U_1V_1(x^-_1-x^+_2)(1-x_1^- x_2^-)\gamma_1\gamma_2}
\nonumber \\
a_8&=&\frac{2i}{g}\frac{U_2V_2\,  x^-_1x^-_2(x^+_1-x^+_2)\gamma_1\gamma_2}{q^{\frac{3}{2}} x^+_1x^+_2(x^-_1-x^+_2)(x^-_1x^-_2-1)}\nonumber \\
a_9&=&\frac{(x^-_1-x^+_1)\gamma_2}{(x^-_1-x^+_2)\gamma_1} \nonumber \\
\nonumber
a_{10}&=&\frac{U_2V_2 (x^-_2-x^+_2)\gamma_1}{U_1V_1(x^-_1-x^+_2)\gamma_2}\, .
\eea
Here the basic variables $x^{\pm}$ parametrizing a fundamental representation of the centrally extended superalgebra $\psu_q(2|2)$ satisfy the following constraint \cite{Beisert:2008tw}
\bea
\label{fc}
\frac{1}{q}\left(x^++\frac{1}{x^+}\right)-q\left(x^-+\frac{1}{x^-}\right)=\left(q-\frac{1}{q}\right)\left(\xi+\frac{1}{\xi}\right)\, ,
\eea
where the parameter $\xi$ is related the coupling constant $g$ as
\bea
\xi=-\frac{i}{2}\frac{g(q-q^{-1})}{\sqrt{1-\frac{g^2}{4}(q-q^{-1})^2}}\, .
\eea
The (squares of) central charges are given by
\bea
U_i^2=\frac{1}{q}\frac{x^+_i+\xi}{x^-_i+\xi}=e^{ip_i}\, , ~~~~V^2_i=q\frac{x^+_i}{x^-_i}\frac{x^-_i+\xi}{x^+_i+\xi} \, ,
\eea
and the parameters $\gamma_i$ are
\bea
\gamma_i=q^{\frac{1}{4}}\sqrt{\frac{ig}{2}(x^-_i-x^+_i)U_iV_i}\, .
\eea
The q-deformed dispersion relation ${\cal E}$ takes the form
\be\la{qdisp}
\Bigg(1-\frac{g^2}{4}(q-q^{-1})^2 \Bigg)\Bigg(\frac{q^{{\cal E}/2}-q^{-{\cal E}/2}}{q-1/q}\Bigg)^2-g^2\sin^2\frac{p}{2}=\Bigg(\frac{q^{1/2}-q^{-1/2}}{q-1/q}\Bigg)^2\, .
\ee
Finally, we point out that in the q-deformed dressing phase the variable $u$ appears which is given by 
\bea
u(x)=\frac{1}{\qdp}\log\Bigg[-\frac{x+\tfrac{1}{x}+\xi+\tfrac{1}{\xi}}{\xi-\tfrac{1}{\xi}}\Bigg]\, .
\eea


\section{Expansion of the q-deformed Gamma-function}\la{app:qGamma}

We take $q=e^{i\qdp/g}$, keep $x$ fixed and send $g\to\infty$. We are interested in the leading term only. At the end we analytically continue to imaginary $\qdp$. We have \cite{Ben}
{\small
\bea\nonumber
 &&\log\frac{\Gamma_{q^2}(1+g x)}{\Gamma_{q^2}(1-gx)} = -i\qdp x + \int_0^\infty {dt\ov t}\Big( \frac{2 \left(e^{-\qdp x t}-e^{\qdp x t}\right)}{\left(e^{\frac{\qdp t}{g}}-1\right)
   \left(e^{\pi t}-1\right)}-\frac{g\pi \left(e^{-\qdp x t}-e^{\qdp x t}\right)}{\qdp 
   \left(e^{\pi t }-1\right)^2}-\frac{\qdp
   \left(e^{-\qdp x t}-e^{\qdp x t}\right)}{g\pi
   \left(e^{\frac{\qdp t}{g}}-1\right)^2}~~~~~~\\\nonumber
 &&+\frac{g\pi \left(e^{-\qdp x t}-e^{\qdp x t}\right)}{\qdp 
   \left(e^{\pi t }-1\right)^2}-\frac{\qdp 
   \left(e^{-\qdp x t}-e^{\qdp x t}\right)}{g\pi
   \left(e^{\frac{\qdp t}{g}}-1\right)^2}+\frac{2 g x
   e^{\frac{\qdp t}{g}}}{e^{\pi t }-1}+\frac{2 g x
   e^{t \left(\pi+\frac{\qdp}{g}\right)}}{e^{\pi t
   }-1}+\frac{e^{-\qdp x t}-e^{\qdp x t}}{e^{\frac{\qdp t}{g}}-1}+\frac{e^{-\qdp x t}-e^{\qdp x t}}{e^{\pi t }-1}\Big)\,.\\\la{lnGovG}
\eea
}
\begin{figure}[t]
\begin{center}
\includegraphics[width=0.36\textwidth]{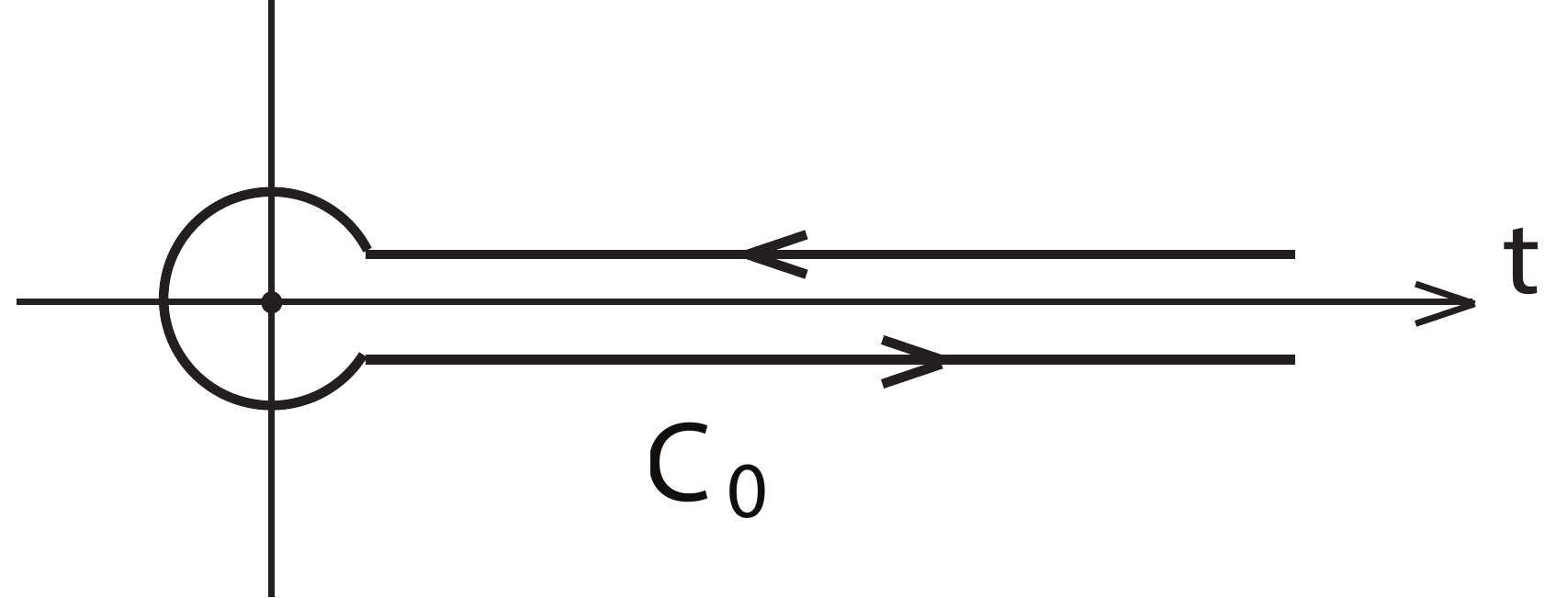}
\caption{\small The integration contour $C_0$ in the integral $\int_{C_0}\, {dt\ov 2\pi i}\, F(t)\, \ln(-t)$.}
\label{C0}
\end{center}
\end{figure}
We understand  integrals of the form $\int_0^\infty dt F(t)$ as in \cite{JKM}\be
\int_{0}^\infty\, {dt}\, F(t) \equiv 
\int_{C_0}\, {dt\ov 2\pi i}\, F(t)\, \ln(-t)\,,
\ee
where the integration contour $C_0$ goes from $+\infty+i0$ above the real axis, then around zero, and finally below the real axis to  $+\infty-i0$, see Figure \ref{C0}. Then the terms on the second line of \eqref{lnGovG} can be easily computed by using the functions introduced in \cite{BF}\footnote{The function $F_2$ is a simple modification of the one introduced in \cite{BF}.}
\bea\nonumber
F_2(z,a)&\equiv&\int_0^\infty\, {dt\ov t}\,{e^{z a t}\ov (e^{a t}-1)^2} =\frac{13}{12}+\frac{z^2}{2}-\frac{3
   z}{2}-\gamma\left(\frac{z^2}{2}-z+\frac{15}{12}\right) +(z-1) \text{log$\Gamma $}(2-z)\\\nonumber
   &&\qquad\qquad\qquad +\psi ^{(-2)}(2-z)-\log (A)-\frac{1}{2} \log (2 \pi )-\left(\frac{z^2}{2}-z+\frac{15}{12}\right) \log a\,,
\\\nonumber
F_1(z,a) &\equiv&\int_0^\infty\, {dt\ov t}\,{e^{z a t}\ov e^{a t}-1} = F_2(z+1,a)-F_2(z,a)\\
&=&
-\gamma ( z-\frac{1}{2})+\text{log$\Gamma $}(1-z) -
z \log a+\frac{1}{2} \log \left(\frac{a}{2 \pi
   }\right)\,,\\\nonumber
F_0(z) &\equiv&\int_0^\infty\, {dt\ov t}\,{e^{z t}} = F_1(z+1,1)-F_1(z,1)=-\g -\log(-z)\,,
\eea
where $\psi ^{(-2)}\left(z\right)$ is given by
\be
\psi ^{(-2)}\left(z\right) = \int_0^z\, dt\,  \text{log$\Gamma
   $}\left(t\right)\,,
\ee
and $A$ is Glaisher's constant which satisfies $\log(A)=1/12-\zeta^\prime(-1)$ and $\zeta$ is the Riemann zeta function. 

Thus, the terms on the second line of \eqref{lnGovG} are equal to 

\bea\nonumber
i_2&=&\frac{g\pi  }{\qdp
   }\left(F_2\left(-\frac{\qdp 
   x}{\pi },\pi \right)-F_2\left(\frac{\qdp 
   x}{\pi },\pi \right)\right)+\frac{\qdp}{g\pi}
\left(F_2\left(-g
   x,\frac{\qdp }{g}\right)-F_2\left(g
   x,\frac{\qdp }{g}\right)\right)   \\\nonumber
   &+&2 g x
   F_1\left(\frac{\qdp }{\pi  g},\pi \right)+2 g
   x F_1\left(1-\frac{\qdp }{\pi  g},\pi
   \right)+F_1\left(-g x,\frac{\qdp }{g}\right)-F_1\left(g
   x,\frac{\qdp }{g}\right)\\\la{secondline}
   &+&F_1\left(-\frac{\qdp  x}{\pi },\pi
   \right)-F_1\left(\frac{\qdp  x}{\pi },\pi
   \right)
\,.
\eea
The integral on the first line of \eqref{lnGovG} is convergent at $t=0$, and one can expand the integrand in powers of $1/g$. One gets for the leading term
\bea
i_1=g\int_0^\infty {dt\ov t}\Big(\frac{2 \sinh (\qdp xt)}{\pi  \qdp 
   t^2}+\frac{4 \sinh
   (\qdp xt)}{\qdp  t(1-e^{\pi t})}+\frac{2 \pi  \sinh (\qdp xt)}{\qdp 
   \left(e^{\pi  t}-1\right)^2}\Big)\,.
\eea
These integrals can be computed by using the functions
\bea
H_0(z)&\equiv&\int_0^\infty\, {dt\ov t}\,{e^{z t}\ov t^2} =-\frac{1}{4} z^2 (2 \log (-z)+2 \gamma -3)\,,\\\nonumber
G_1(z,a)&\equiv&\int_0^\infty\, {dt\ov t}\,{e^{z a t}\ov a t(e^{a t}-1)} =\log
   (A)+\left(-\frac{z^2}{2}+\frac{z}{2}-\frac{1}
   {12}\right) \log (a)+\gamma 
   \left(-\frac{z^2}{2}+\frac{z}{2}-\frac{1}{12}
   \right)\\
   &&-\psi
   ^{(-2)}(1-z)+\frac{1}{2}(1-z) \log (2 \pi )\,.
\eea
Summing up $i_1$ and $i_2$ and taking the large $g$ limit one gets for  the leading term
\bea\la{qGamma1}
\log\frac{\Gamma_{q^2}(1+g x)}{\Gamma_{q^2}(1-gx)} &\approx& g\Big(-2 x+2 x \log (g)+x \big(\log (-x)+  \log
   (x)\big)\Big)\\\nonumber
&+&   g{2\pi\ov \qdp} \big(\psi ^{(-2)}(1-{\qdp x\ov\pi} )-\psi^{(-2)}(1+ {\qdp x\ov\pi} )\big)\,.
\eea
The analytic continuation of this expression to imaginary $x$ and $\qdp$ is straightforward.


\end{document}